\documentstyle[12pt]{article}
\newcommand{\be}{\begin{eqnarray}}
\newcommand{\ee}{\end{eqnarray}}
\newcommand{\ba}{\begin{array}}
\newcommand{\ea}{\end{array}}
\newcommand{\half}{{\textstyle{\frac{1}{2}}}}
\newcommand{\textfrac}[2]{{\textstyle{\frac{#1}{#2}}}}
\newcommand{\fourint}[1]{\int\!\frac{d^4 #1}{(2\pi)^4}}
\newcommand{\Fdual}{\widetilde{F}}

\newcommand{\cU}{{\cal U}}
\newcommand{\tr}{{\rm tr}\,}

\newcommand{\kslash}{k\hspace{-.5em}/\hspace{.15em}}

\newcommand{\effop}[1]{\mbox{``$#1$''}}
\begin{document}
%
%
\rightline{RUB-TPII-7/97}
\rightline{hep-ph/9710534}
\vspace{.3cm}
\begin{center}
\begin{large}
{\bf Estimates of higher--dimensional vacuum condensates from
the instanton vacuum} \\
\end{large}
\vspace{1.4cm}
{\bf M.V. Polyakov}$^{\rm 1,2}$, {\bf C. Weiss}$^{\rm 2}$ \\
\vspace{0.2cm}
$^{\rm 1}${\em Theory Division of Petersburg Nuclear Physics Institute \\
188350 Gatchina, Leningrad District, Russian Federation}\\
\vspace{0.2cm}
$^{\rm 2}${\em Institut f\"ur Theoretische Physik II \\
Ruhr--Universit\"at Bochum \\
D--44780 Bochum, Germany}
\end{center}
\vspace{1cm}
\begin{abstract}
\noindent
We calculate the values of non-factorizable dimension--7 vacuum condensates
in the instanton vacuum. We comment on a method, recently proposed by
Oganesian, to estimate the dimension--7 condensates by factorization
of dimension--10 condensates in various inequivalent ways.
The instanton estimates could be used to analyze the stability
of QCD sum rules with increasing dimensions.
\end{abstract}
\vspace{1cm}
\newpage
In applications of the QCD sum rule method one frequently requires
the vacuum expectation values of operators of high dimension.
One such instance are power corrections to
the moments of polarized structure functions (Bjorken and Ellis--Jaffe
sum rules), which were originally computed including operators up to
dimension 8 \cite{BBK90,Ioffe};
later the calculations were extended to
dimension 10 \cite{Stein95,Oganesian}
(see ref.\cite{Ioffe1} for a critical analysis of the QCD sum rule
calculations). The usual way to estimate the values of the
higher--dimensional condensates is by factorization,
{\em i.e.}, if possible one writes the operator as a product of
lower--dimensional operators with non-vanishing VEV and assumes dominance
of the vacuum intermediate state. This approximation is justified
in the large--$N_c$ limit.
\par
Recently, an interesting extension of the factorization method was
proposed by Oganesian \cite{Oganesian}. His method is based on the
observation that some high--dimensional operators can be factorized in
different ways by making use of the QCD equations of motion. Assuming
equivalence of the different factorized approximations to the
high--dimensional condensate
(``self--consistent factorization'') he obtains relations between the
lower--dimensional condensates arising in the factorization. In this way,
starting from a dimension--10 operator, he derives a set of relations between
dimension--7 operators which, together with some additional assumptions,
allow him to estimate the values of the dimension--7 condensates. This,
in turn, furnishes a factorized approximation to the original dimension--10
condensate.
\par
The aim of this note is to address two points. First, we would like to
comment on the approach of ref.\cite{Oganesian} from the point of
view of the $1/N_c$--expansion.
We point out that not all of the relations obtained there
are consistent with the $1/N_c$--expansion.
Second, for a rough quantitative
check we estimate the vacuum expectation values of the dimension--7
operators in the instanton vacuum. We use a method by which QCD
operators are systematically represented as effective operators
in the effective chiral theory derived from the instanton
vacuum \cite{DPW96,PW96}.
While the values of higher--dimensional condensates in the instanton
vacuum have no physical significance by themselves, the estimates can
nevertheless be used to assess the convergence of QCD sum rule
calculations with increasing dimension.
\par
In ref.\cite{Oganesian} an attempt is made to determine the vacuum 
expectation values of the following dimension--7 operators, which 
themselves are not factorizable:
\be
R_d &=& \frac{1}{12} \langle \, \bar \psi \frac{\lambda^a}{2} \psi 
\; d^{abc} F^b_{\mu\nu} F^c_{\mu\nu} \, \rangle ,
\label{R_d}
\\
R_f &=& \frac{1}{12} \langle \, \bar \psi \frac{\lambda^a}{2} 
\sigma_{\alpha\beta} \psi \; f^{abc} F^b_{\mu\alpha} F^c_{\mu\beta} \, 
\rangle ,
\label{R_f}
\\
S_1 &=& \frac{1}{24} \langle \, \bar \psi i \gamma_5 \psi \;
F^a_{\mu\nu} \Fdual^a_{\mu\nu} \, \rangle ,
\label{S_1}
\\
S_d & = & \frac{1}{12} \langle \, \bar \psi \frac{\lambda^a}{2} \gamma_5
 \psi \; d^{abc} F^b_{\mu\nu} \Fdual^c_{\mu\nu} \,
\rangle ,
\label{S_d}
\ee
where we have set $F_{\mu\nu} \equiv g G_{\mu\nu}$.
Applying the ``self-consistent factorization'' described above
the following relations between dimension--7 condensates were derived
in ref.~\cite{Oganesian}:

\be
\frac 16 R_1+\frac 12 R_d
+R_f +\frac 37 N &\approx& 0\;,
\label{bad1}
 \\
R_d+R_f +\frac{22}{21}  N &\approx& 0\;,
\label{good1}  \\
S_d-\frac{4}{21}  N &\approx& 0\;,
\label{good2}
\ee
where we have introduced the notation
\be
R_1 &=& \frac{1}{24} \langle F_{\mu\nu}^a F_{\mu\nu}^a\rangle
\langle \bar \psi \psi\rangle , \\
N &=& \frac{1}{24}
\langle \bar \psi\frac{\lambda^a}{2} 
\sigma_{\alpha\beta} \psi \, F_{\alpha\beta}^a \rangle^2 \, / \,
\langle \bar \psi \psi\rangle .
\ee
Let us note that in the course of the derivation of
Eqs.(\ref{bad1}--\ref{good2})
in ref.~\cite{Oganesian} it was assumed that $S_1\approx S_d\sim 0$, and some
of the four--fermionic condensates were neglected.
We would like to comment on the assumption that $S_d\approx S_1\sim 0$. 
The first relation follows from Eq.(\ref{good2}), which implies 
$S_d\approx N/5$,
which is zero at the given level of accuracy ($\sim 30\%$). We shall see
below that indeed $S_d\approx N/5$ with the values obtained from the 
instanton vacuum. The smallness of $S_1$, on the other hand, was obtained
in ref.~\cite{Oganesian} by representing $S_1$ as a difference of two
factorizable vacuum averages,
\be
S_1 &=& \frac{1}{24} \left( \langle \, \bar\psi\psi \; F^2 \, \rangle
-\frac{1}{2}
\langle \, \bar \psi\sigma_{\alpha\beta} \sigma_{\gamma\delta}\psi \; 
F^a_{\alpha\beta} F^a_{\gamma\delta} \, \rangle
\right) ,
\ee
which were argued to cancel each other after factorization. However, the
large--$N_c$ analysis shows that this cancellation occurs at level $N_c^2$,
whereas $S_1$ itself is of order $N_c$. Thus the cancellation of the
$O(N_c^2)$ terms is merely a statement about the large--$N_c$ behavior
of $S_1$, but does not imply the smallness of $S_1$ relative to, say,
$N$.
\par
We now want to estimate the values of the dimension--7 condensates in the
instanton vacuum. The basis of our description is the medium of independent
instantons with an effective size distribution, which was obtained by
Diakonov and Petrov as a variational approximation to the interacting
instanton partition function \cite{DP84_1}. The average instanton size
is
\be
\bar\rho &\simeq& (600 \, {\rm MeV})^{-1} .
\ee
Fermions interact with the instantons mainly through the zero modes
associated with the individual instantons. By integrating over the
coordinates of the (anti--) instantons in the medium one derives in
the large--$N_c$ limit an
effective fermion theory of the form of a Nambu--Jona-Lasinio model
with many--fermionic interactions, which describes the dynamical
breaking of chiral symmetry \cite{DP86_prep}. In ref.\cite{DPW96}
a method has been developed by which QCD operators involving gluon fields
can systematically be represented as many--fermion operators. The
averages of these ``fermionized'' operators in the vacuum of the effective
fermion theory can be computed using standard techniques.
It was shown that this method gives matrix elements
fully consistent with the QCD anomalies. The method has been applied
to compute the dimension--5 mixed quark--gluon condensate \cite{PW96}.
Recently, nucleon matrix elements of QCD operators of twist--3 and 4
have been calculated within the description of the nucleon
as a chiral soliton of the effective theory \cite{BPW97}.
\par
Let us consider first the operator of Eq.(\ref{R_f}),
\be
O_{R_f} (x) &=& \frac{1}{12} \; \bar \psi (x) \frac{\lambda^a}{2} 
\sigma_{\alpha\beta} \psi (x) \; f^{abc} F^b_{\mu\alpha} (x) 
F^c_{\mu\beta} (x) .
\label{O_R_f}
\ee
Following \cite{DPW96},
after passing to the Euclidean theory we want to replace the gluonic part 
of this operator, $f^{abc} F_{\mu\alpha}^b (x) F_{\mu\beta}^c (x)$,
by an effective fermion operator. The field strength of one
(anti--) instanton in singular gauge, centered at the origin, is given by
\be
F_{\mu\nu}^a (x)_{I(\bar I)} 
&=& \left[ (\eta^\mp )^a_{\rho\nu} \frac{x_\mu x_\rho}{x^2}
+ (\eta^\mp )^a_{\mu\rho} \frac{x_\rho x_\nu}{x^2}
- \frac{1}{2} (\eta^\mp )^a_{\mu\nu} \right]
\frac{8 \rho^2}{(x^2 + \rho^2)^2} ,
\ee
where $(\eta^\pm )^a_{\rho\nu} = \eta^a_{\rho\nu},
\bar\eta^a_{\rho\nu}$ are the 't~Hooft symbols; the field takes values
in the algebra of the group $SU(2)$. The instanton is embedded
in the color group $SU(N_c)$ by identifying the $SU(2)$ group with
the subgroup spanned by the generators $\lambda^1, \lambda^2, \lambda^3$.
For a general instanton, with color orientation described by an $SU(N_c)$
matrix, $\cU$, and centered at $z$, one has
\be
\lefteqn{
\left( f^{abc} F_{\mu\alpha}^b (x) F_{\mu\beta}^c (x) \right)_{I(\bar I)} }
&& \nonumber \\
&=& - \half \tr [\lambda^a \cU \lambda^b \cU^\dagger ] \;
\left[ (\eta^\mp )^b_{\rho\alpha} \frac{y_\rho y_\beta}{y^2}
+ (\eta^\mp )^b_{\alpha\rho} \frac{y_\rho y_\beta}{y^2}
- \frac{1}{2} (\eta^\mp )^b_{\alpha\beta} \right]
\frac{64 \rho^4}{(y^2 + \rho^2)^4} ,
\label{dFF_U}
\ee
where $y = x-z$.
In zero mode approximation \cite{DP86}, the interaction of the quark field
($\psi^\dagger \equiv i \bar\psi$) with the (anti--) instanton with 
collective coordinates $\cU, z$ and size $\rho$ is given by
\be
V_\pm [\psi^\dagger , \psi ]
&=& 4\pi^2\rho^2
\fourint{k_1}\fourint{k_2} \, e^{i z\cdot (k_2 - k_1 )} \;
F(k_1 ) F(k_2 ) \nonumber \\
&& \times \; \psi^\dagger (k_1 ) \;
\left[ \cU \tau^\mp_\mu \tau^\pm_\nu \cU^\dagger \right]
\;
\left[ \frac{1}{8}
\gamma_\mu \gamma_\nu \frac{1 \pm \gamma_5}{2} \right] \;
\psi (k_2 ) .
\label{V_I}
\ee
Here $F(k)$ is a form factor of width $\rho^{-1}$, proportional to the
Fourier transform of the wave function of the fermion zero mode, with
$F(0) = 1$, and $\tau^\pm_\kappa$ are $N_c \times N_c$ matrices with
$(\tau, \mp i)$ in the upper left corner and zero
elsewhere \cite{DP86}.
The many--fermion vertex corresponding to
$f^{abc} F_{\mu\alpha}^b (x) F_{\mu\beta}^c (x)$
is then defined as the average of the product of Eq.(\ref{dFF_U}) with
Eq.(\ref{V_I}) over the collective coordinates of one
instanton \cite{DPW96},
\be
(Y_{fFF\pm})^a_{\alpha\beta}(x)[\psi^\dagger , \psi ]
&=& i \frac{N_c M}{4\pi^2\bar\rho^2} \int d^4 z \int d\cU \;
\left( f^{abc} F_{\mu\alpha}^b (x) F_{\mu\beta}^c (x) \right)_{I(\bar I)}
\; V_\pm [\psi^\dagger , \psi ] .
\nonumber \\
\label{Y_ddF_1}
\ee
Performing the integral over color orientations in the leading order
of the $1/N_c$--expansion, Eq.(\ref{Y_ddF_1}) becomes
\be
(Y_{fFF\pm})^a_{\alpha\beta}(x)[\psi^\dagger , \psi ]
&=& \frac{i M \bar\rho^2}{N_c}
\fourint{k_1}\fourint{k_2} \; e^{i k\cdot x } \; {\cal G}_{fFF}(k) \;
F(k_1 ) F(k_2 ) \nonumber \\
&& \times \; \psi^\dagger (k_1 ) \; \frac{\lambda^a}{2} \;
\Gamma_{\alpha\beta} \frac{1 \pm \gamma_5}{2} \psi (k_2 ) ,
\nonumber \\
\Gamma_{\alpha\beta} &=& \sigma_{\rho\beta} \frac{k_\alpha k_\rho}{k^2}
+ \sigma_{\alpha\rho} \frac{k_\rho k_\beta}{k^2}
- \half \sigma_{\alpha\beta}, \hspace{1.5cm}
k \; = \; k_2 - k_1 . \label{Y_fFF}
\ee
Here we have introduced the Fourier transform of the function
$f^{abc} F_{\mu\alpha}^b F_{\mu\beta}^c$ of the (anti--) instanton
field,
\be
\lefteqn{ \int d^4 x \; \left( f^{abc} F_{\mu\alpha}^b (x) F_{\nu\beta}^c (x)
\right)_{I (\bar I)} \exp (-ik\cdot x) } && \nonumber \\
&=& {\cal G}_{fFF} (k)
\left[ (\eta^\mp )^a_{\rho\beta} \frac{k_\alpha k_\rho}{k^2}
\; + \; (\eta^\mp )^a_{\alpha\rho} \frac{k_\rho k_\beta}{k^2}
\; - \; \half (\eta^\mp )^a_{\alpha\beta} \right] ,
\hspace{.5cm}
\label{F_fourier}
\ee
\be
{\cal G}_{fFF}(k) &=& - 64 \rho^4 \int d^4 x \frac{1}{(x^2 + \rho^2 )^4}
\left[ 1 + \frac{4}{3} \left(\frac{(k\cdot x)^2}{k^2 x^2} - 1\right)
\right] \exp (-ik\cdot x) \nonumber \\
&=& 32 \pi^2 \left\{ 
-\left( \frac{t^2}{6} + 4 + \frac{32}{t^2} \right) K_0 (t)
-\left( t + \frac{16}{t} + \frac{64}{t^3}\right) K_1 (t)
+ \frac{64}{t^4}
\right\} ,
\nonumber \\
&& t \; = \; k\rho ,
\label{G_3}
\ee
{\em cf.}\ the expression for the Fourier transform of the instanton field
in ref.\cite{PW96}.
\par
With $f^{abc} F_{\mu\alpha}^a (x) F_{\mu\beta}^b (x)$ replaced by
the many--fermion vertices, Eq.(\ref{Y_fFF}), resulting from
$I$'s and $\bar I$'s, the operator Eq.(\ref{O_R_f})
is represented by the effective quark operator
\be
\effop{O_{R_f}}(x)[\psi^\dagger , \psi ] &=&
\frac{-i}{12} \, \psi^\dagger (x) \frac{\lambda^a}{2} \sigma_{\alpha\beta} 
\psi (x) \; \left[ (Y_{fFF+})^a_{\alpha\beta} (x) \; 
+ \; (Y_{fFF-})^a_{\alpha\beta} (x) \right] \;.
\label{effop}
\ee
Computing the average of Eq.(\ref{effop}) in the vacuum of the
effective fermion theory derived from the instanton medium,
with the fermion propagator given by
\be
G(k) &=& (\kslash - i M F^2 (k))^{-1},
\ee
where $M$ is the dynamically generated quark mass at zero momentum,
\be
M &\simeq& 350 \, {\rm MeV},
\ee
we find
\be
R_f &=& \langle \effop{O_{R_f}} \rangle_{\rm eff} \;\; = \;\;
\frac{N_c M}{24\pi^2\, \bar\rho^6} \, I^{(2)}(M\bar\rho ) ,
\ee
\be
I^{(2)}(M \bar\rho ) &\equiv&
4 \pi^2 \bar\rho^6 \fourint{k_1}\fourint{k_2} \;
\frac{{\cal G}_{fFF}(k) \, F(k_1 ) \, F(k_2 ) N(k_1 , k_2 )}
{\left[ k_1^2 + M^2 F^4 (k_1) \right]
\left[ k_2^2 + M^2 F^4 (k_2) \right]} ,
\label{I2} \\
N(k_1 , k_2 ) &=& \textfrac{1}{4}
\tr [\sigma_{\alpha\beta} (\kslash_1 + i M(k_1 ))
\Gamma_{\alpha\beta} (\kslash_2 + i M(k_2 )) ] \nonumber \\
&=& \frac{1}{k^2} \left[ 8 k_1^2 k_2^2 - 6 (k_1^2 + k_2^2 )
k_1\cdot k_2 + 4 (k_1\cdot k_2 )^2 \right] . \nonumber
\ee
Evaluating the integral numerically we find
\be
I^{(2)} (M\bar\rho = 0) &=& \frac{64}{5} .
\ee
(The limit $M\bar\rho \rightarrow 0$ corresponds to working in the leading
order of the packing fraction of the instanton medium). Thus,
\be
R_f &=&\frac{16}{15} \times \frac{N_cM}{2\pi^2\bar\rho^6}
\label{R_fnum}
\ee
Along the same lines we can compute the values of the other 
condensates, Eqs.(\ref{R_d}, \ref{S_1}, \ref{S_d}). The effective fermion
vertex for the function of the gauge field involving the totally symmetric
structure constants, $d^{abc}$, in Eqs.(\ref{R_d}, \ref{S_d}) can easily be 
determined in analogy to Eq.(\ref{dFF_U}). One finds 
\be
\left.
\begin{array}{lcr}
R_d &=& {\displaystyle  \frac{8}{45}} \\[.5cm]
S_1 &=& {\displaystyle -\frac{8}{15}} \\[.5cm]
S_d &=& {\displaystyle -\frac{8}{45}} 
\end{array}
\right\} \times \frac{N_cM}{2\pi^2\bar\rho^6}
\label{R_dS_dnum}
\ee
The values of the gluon, quark, and mixed quark--gluon condensates
in the instanton vacuum are as follows \cite{DPW96,PW96}:
\be
\langle F_{\mu\nu}^a F_{\mu\nu}^a
\rangle &=&  16 \, C \, \frac{N_c M^2}{\bar\rho^2},
\hspace{2cm} C \; \approx \; 0.335,\\
\langle \bar \psi \psi \rangle &=& -\frac{N_cM}{2\pi^2\bar\rho^2}\;,\\
\langle \bar \psi\frac{\lambda^a}{2} \sigma_{\alpha\beta} \psi
\; F_{\alpha\beta}^a \rangle
&=& -4 \times \frac{N_cM}{2\pi^2\bar\rho^4}\;.
\ee
(In order to express the gluon condensate in terms of the dynamical quark 
mass we have made use of the self--consistency condition defining the
dynamical quark mass \cite{DP86,DPW96}.)
Having all these values at hand we are now able to check the relations
Eqs.(\ref{bad1}--\ref{good2}). Grouping the contributions with
different signs on different sides one gets, in units of $N_c
M/2\pi^2\bar\rho^6$, 
\be
1.16 \approx 0.4 \; ,
\label{bad1_num}
\\
1.24 \approx 0.7 \;,
\label{good1_num}
\\
0.17 \approx 0.13 \;.
\label{good2_num}
\ee
We see that the first relation Eq.(\ref{bad1})
works poorly numerically,
whereas the other relations are satisfied with reasonable accuracy.
Probably the failure of Eq.(\ref{bad1})  is due to the neglection
in ref.~\cite{Oganesian} of $S_1$ contributions.
In our estimate $S_1\approx 0.8 N$ is not negligible, contrary
to the claim of ref.\cite{Oganesian}, which is based on inconsistent
large--$N_c$ counting.
\par
To summarize, we have estimated the values of the dimension--7
non-factorizable quark--gluon condensates in the instanton model of the
QCD vacuum; the estimates are given by
Eqs.(\ref{R_fnum}, \ref{R_dS_dnum}). In contrast to the estimates based
on the factorization method of ref.\cite{Oganesian}, namely
$S_1 \approx S_d \approx R_f \approx 0$ and $R_d \approx -R_1$,
we obtain rather $S_d \approx R_d \approx 0$ and 
$-R_f \approx S_1 \approx N$, in the sense of 50\% accuracy.
\par
We stress that individual values of the higher--dimensional condensates
have no direct physical significance. Nevertheless, it would be extremely
interesting to reanalyze the stability of QCD sum rules using the set of
values of high dimensional condensates estimated in the instanton vacuum.
\par
Finally, we note that it is possible to evaluate the corrections to
the Bjorken and Ellis--Jaffe sum rules in the instanton vacuum
directly \cite{BPW97}, not relying on an expansion of correlation functions
in terms of local condensates of increasing dimension as in the QCD sum 
rule method. The twist--3 and 4 flavor nonsinglet corrections, 
$d^{(2)}_{NS}$ and $f^{(2)}_{NS}$, have been computed in this approach
in ref.\cite{BPW97} and are in reasonable agreement with the sum rule
results of refs.\cite{BBK90,Ioffe,Stein95}.
\\[1cm]
{\large\bf Acknowledgements} \\[.3cm]
We are grateful to A.G.~Oganesian for valuable discussions and 
correspondence. We wish to thank Klaus Goeke for encouragement and 
multiple help.
\\[.2cm]
This work has been supported in part by the Deutsche Forschungsgemeinschaft,
by the Russian Foundation for Basic Research, and by COSY (J\"ulich).
M.V.P.\ is supported by the A.v.Humboldt Foundation.
%
%

\end{document}